# Realização doméstica e escolar de foto e vídeo 3D

José J. Lunazzi

Instituto de Física-Universidade Estadual de Campinas-UNICAMP, Rua Sergio Buarque de Holanda, 777 Campinas, 13083-859-São Paulo-SP, Brazil
lunazzi at ifi.unicamp.br

**Resumo**
A indústria não tem dado atenção ao mercado doméstico ou profissional de pequena escala nas novas tecnologias digitais para imagem 3D. O que foi feito ao longo do século XX usando filme fotográfico não está hoje ao alcance das pessoas, sendo que a facilidade de uso é muito maior. Técnicas desenvolvidas no Brasil há mais de vinte anos, e o uso de simples câmeras convencionais de fotografia e vídeo permítem realizar fotos e filmes caseiros por meio de conversões digitais na edição e o uso de óculos bicolor. A técnica anaglífica, a mesma que a NASA usa para mostrar ao público as imagens de Marte, por exemplo, não teve espaço no Brasil ainda. Vamos analisar os motivos que podem estar influenciando e os caminhos para mudar isso.

**Introdução**
A estereoscopia no Brasil vem sendo aplicada desde o século XIX, existem até fotos estéreo do ano 1865 e do Imperador Dom Pedro II(1)(2). Mas é bem provável que aquelas fotos tenham sido realizadas com câmeras estéreo (com dupla objetiva e sincronismo de disparo para a tomada dupla) importadas. Ou seja, com o mais moderno da tecnologia da época. O mesmo poderiamos dizer do acontecido no século XX, sendo que temos referências até o ano 1939(3) porém muito pouco se encontra acontecendo depois. A utilização da técnica no Brasil é reportadamente muito limitada(4)(5)(6). No meu conhecimento, até 1988 ainda fazia-se em São Paulo fotografias estéreo para, por exemplo, casamentos, em diapositivas coloridas montadas aos pares para introduzir em um visor de duas lentes(4). A técnica parece ter sido esquecida até que Lunazzi em 1987 adapta uma câmera de vídeo para fazer filmação estéreo(7) projetada em uma tela holográfica que desenvolveu para que pudesse ser utilizada por primeira vez com luz branca, onde pode ser vista sem precisar de óculos(8). Foi gerado em 1989 o primeiro filme de longa metragem em vídeo 3D, cinquenta minutos de mímica atoral(9). Em 1993 Lunazzi realiza algumas filmações curtas sobrepondo objetos a projeções de maneira a filmar com uma única câmera que registra a repetição da cena também desde um segundo ponto de vista(10). Segue depois a realização de breves cenas de desenho e vídeo estereoscópico obtidas por meio de computador(11) e que agora podiam ser vistas na tela do computador porque era colorida e com separação nas cores básicas ("RGB"), algo que a TV na época não tinha.

Podemos citar o aparecimento nos anos 90 de algumas fotografias e desenhos publicadas em revistas e jornais(12), que acompanhavam um jogo de óculos bicolor ("anaglifos"). Podia-se notar que no caso dos jornais não houve o devido cuidado no uso das tintas, que geravam a visão das imagens fantasmas indesejadas quando deviam isolar somente uma do par estéreo para cada olho, e podemos atribuir isto à falta de uma cultura técnica que respeite os padrões da estereoscopia. Somente em 2010 a técnica é retomada quando a introdução do 3D no cinema digital trouxe novamente o interesse geral pela estereoscopia. Lunazzi(7) retoma a técnica de aproveitamento de uma câmera só acrescida agora das técnicas de edição digital que facilitam a composição e montagem. É essa facilidade, a de registrar fotografias coloridas instantaneamente e poder ajustar posição e tamanho para melhorar a combinação estereoscópica que descrevo neste artigo, destacando seu uso didático, e também prático, porque ainda não temos no mercado do Brasil câmeras digitais estereoscópicas a venda(13). Em trabalhos mais recentes, fotografias estéreo de pessoas, crianças, esculturas, etc. foram por mim produzidas(14). Uma outra referência é respeito de shows de rock vindos do exterior, que se apresentaram no pais usando projeção estéreo(15)(16), distribuindo óculos mas sem nenhuma oportunidade de o público poder saber o que fazer com eles depois. Algo que poderiamos ter conseguido talvez se tivesse havido iniciativa por parte de produtores locais(17).

Nem tudo é deixado de lado no pais, no entanto: com muita inteligência comercial um microcomputador é vendido como "3D" porque fornece um óculos bicolor para ver anaglifos(18) e um programa para 3D que é um conversor de vídeos 2D a 3D. A técnica de ter duas perspectivas e codificá-las em azul e vermelho, que para alguns preconceituosos poderia parecer uma simples brincadeira, mostrou seu potencial em trabalho recente que permitiu o levantamento amplo e automático da movilidade de espermatozoides(19). Neste trabalho iluminou- se com um diodo vermelho e com um verde, desde duas posições angularmente diferentes para obter a informação em 3D.

**A visão binocular, porque a fotografia não pode recria-la?**

Pode-se achar que tirando uma boa fotografia com a câmera na posição de cada olho e tendo um sistema para oferecer cada tomada a cada olho separadamente, a cena deveria ser vista de maneira perfeita e naturalmente tridimensional. Não é assim e resulta fácil provar: nunca nosso olho focaliza tudo que tem na frente, nem sequer respeitando os 30 cm mínimos que uma adulto tem como limite para focalização próxima. Focaliza-se aos 30 cm deixando de focalizar o que está bem mais longe que isso. Ou, se focalizarmos longe, não focalizamos o que temos a 30 cm. Mas não é somente isto que prova a impossibilidade da estereoscopia resultar natural: os dois olhos convergem sobre um objeto a 30 cm, por exemplo, e tem aos objetos por tras muito separados lateralmente respeito da posição em que aparece o objeto próximo. O que está por tras, de fato, se vê como uma imagem dupla e desfocada, apenas que não considerada pelo cérebro. Igualmente, se observamos um objeto afastado, objetos próximos aparecem desfocados e tão separados de nosso objeto principal que temos imagem dupla novamente. Devemos ter isto em conta antes de começar a fazer nossas fotografias estéreo.

**Tomada do par estéreo com uma câmera só**

A câmera deve estar, sim, em duas posições que possa ser equivalentes à de um olho esquerdo e à de um olho direito de um observador hipotético. Mas a cena deve permanecer estática enquanto a câmera passa de uma posição à outra. A maneira mais simples de começar é com uma cena perfeitamente estática (uma "natureza morta", por exemplo) localizada com seus elementos principais a uma distância média, nem muito próxima nem muito afastada. Uns 3-4m, por exemplo (não irei considerar neste artigo o caso de objetos próximos, a macrofotografia). Colocamos a câmera sobre uma superfície plana, como um banco alto, de onde podemos fazer o enquadramento da cena com somente o giro da câmera na horizontal. Colocamos por tras da câmera uma borda plana como apoio e referência. Um pedaço de madeira como de um sarrafo, por exemplo, bem afixado por peso ou fita adesiva. Centralizamos um objeto como referência da cena e realizamos uma fotografia. Ela vai representar a vista esquerda. Deslocamos a câmera 6,5 cm para a direita (se tudo está bem fixo, nem precisariamos controlar a centralização da referência) e realizamos a tomada que representa a vista direita.

Antes de continuar, explico as vantagens edesta primeira prática: O fato de os objetos de interesse não estarem muito próximos facilita o resultado: podemos notar que dessa maneira não há muita diferença entre centralizar perfeitamente o objeto principal por um ajuste angular da câmera ou deslocar ela paralelamente. Deslocar paralelamente, embora pareça menos natural, gera menos separação entre pontos correspondentes das duas vistas, reduzindo a possibilidade de se chegar a um valor onde o cérebro não consegue fundir as duas imagens do par estéreo em uma só. E, sobretudo, o fato de termos uma base de referência evita a rotação da câmera ao redor do que seria o eixo óptico da lente, um eixo horizontal. Essa rotação é a mais complicada de compensar depois na edição.

**Edição do par estéreo para montagem anaglífica (bicolor)**

Devemos para isto usar um programa de edição de fotografias que permita a eliminação de componentes de tricromia. A composição de cor por meio de vermelho, verde e azul tem de poder ter uma ou duas das componentes eliminadas. Nossa recomendação vai, naturalmente, para um programa livre que pode ser obtido pela internet de graça e para qualquer sistema operacional: o Gimp(20). Abrimos com ele a primeira fotografia e retiramos os canais de cor verde e azul, com o que fica completamente vermelha. Os comandos seriam "Cor" e "Níveis", escolhendo primeiramente "Verde" para colocar o valor zero onde está 255, e o mesmo para "Azul". Além da cor resultante ser fortemente vermelha, pode checar com seu óculos que nada passa pelo olho direito, vé somente preto. Salva a fotografia filtrada acrescentando a letra "E" (de esquerdo) e abre a segunda tomada, para tirar dela as cores vermelho e verde, deixando somente o azul. Confere com o óculos olhando o preto que deve ver pelo olho esquerdo, e salva acrescentando ao nome do arquivo a letra "D" (direito). Vemos na figura 1 o exemplo, tendo na parte de cima as duas tomadas, esquerda e direita respectivamente, e embaixo elas filtradas nos canais de cor.

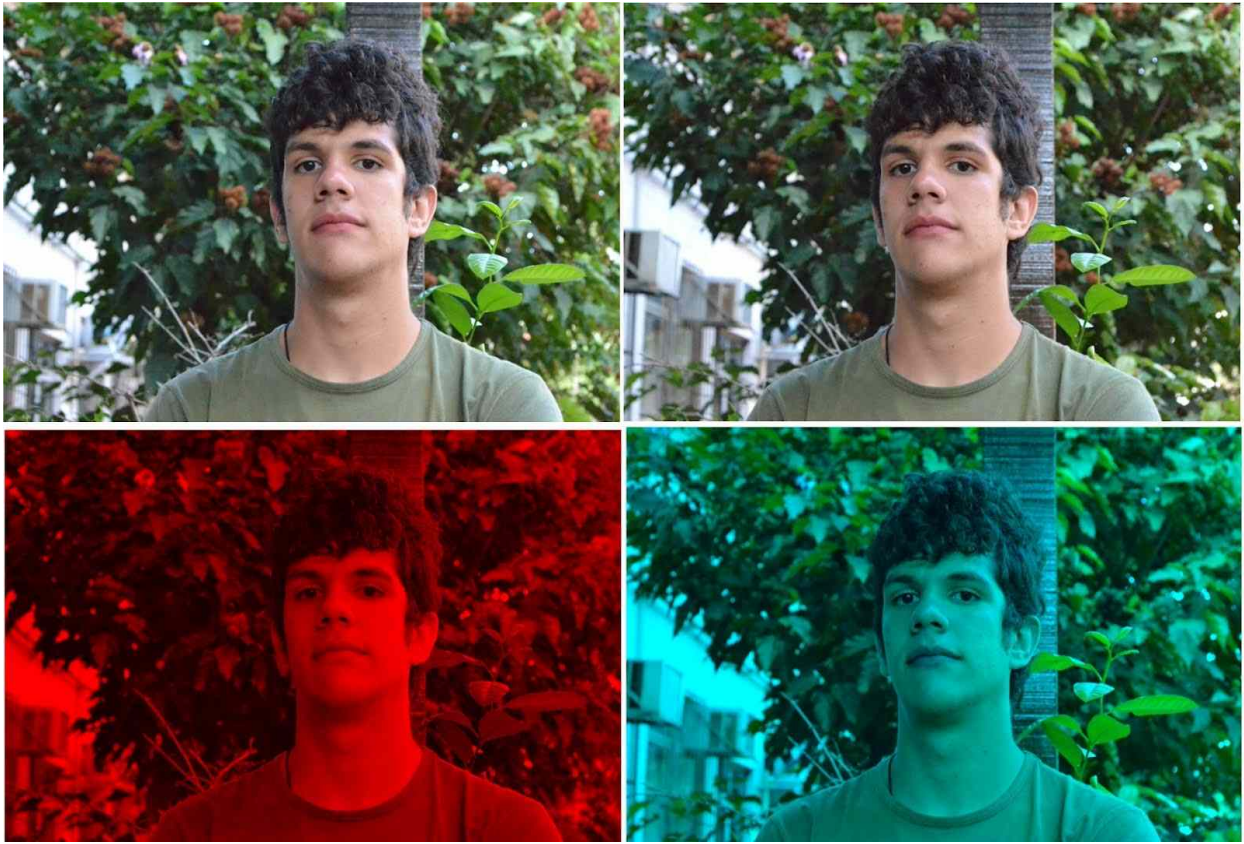

*Figura 1: Acima, tomadas esquerda e direita. Embaixo, as mesmas filtradas.*

Agora parte do menú do GIMP para abrir as duas fotografias juntas usando o comando "Abrir como camadas", o que mostrar somente uma das duas pois a outra fica oculta por baixo. Deve procurar o menú de ferramentas de camadas, que se não apareceu diretamente encontra com CTRL L, e pedir o modo "Adição". Neste modo, os pixels das duas fotos são vissíveis ao mesmo tempo e vemos uma imagem dupla, com contornos vermelho e azul. Se tudo aconteceu como devido, teremos já a visão em 3D quando olharmos pelos óculos, e podemos salvar a foto composta como uma só, em qualquer formato, por exemplo, no .png.

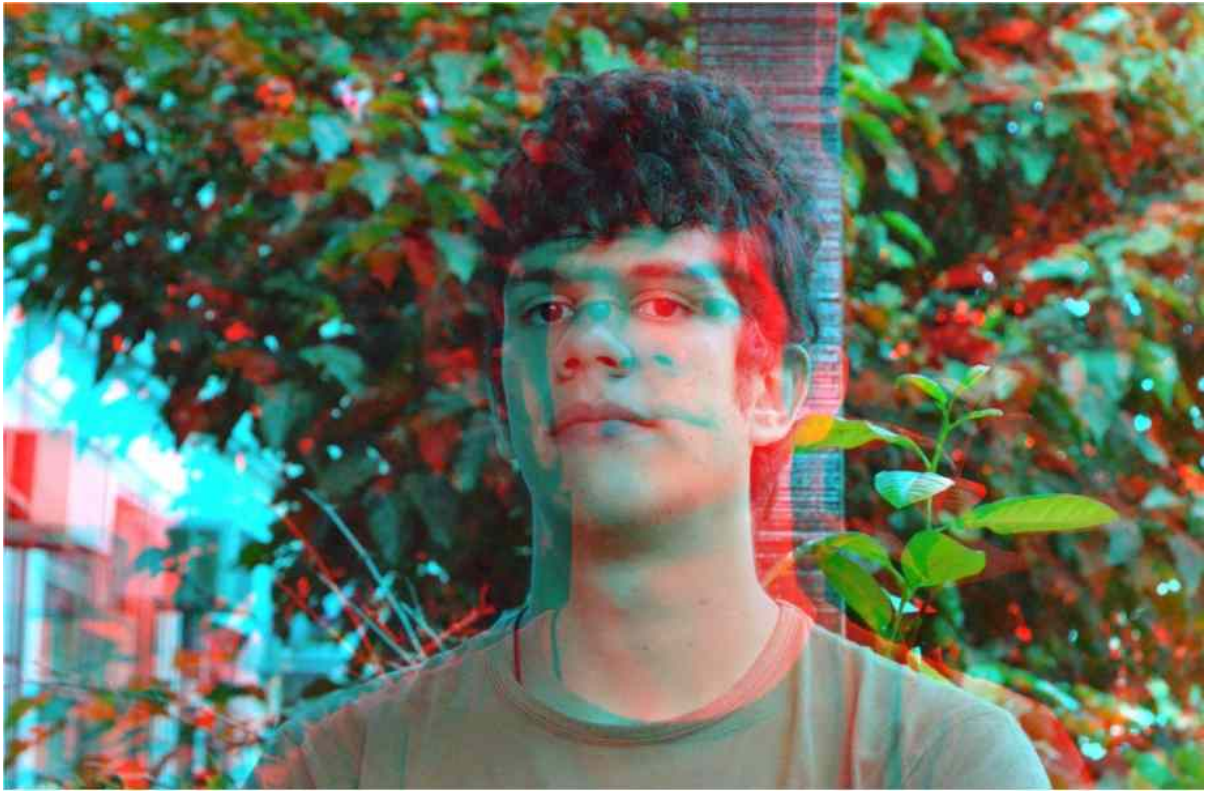

*Figura 2: A fotografia anaglifa resultante.*

Se houve um deslocamento angular porque a câmera não ficou bem apoiada no elemento traseiro, notaremos muito esforço visual para ter a imagem 3D e, sem os óculos, uma separação lateral exagerada dos objetos na cena. Quanto é uma separação "exagerada" é uma questão de prática, pode-se comparar com alguma foto já reconhecidamente boa para ter uma referência. Ou então podemos usar o cursor no comando de deslocamento e fazer com que o elemento principal da cena tenha sobreposição completa. O Gimp permite fazer isto, na fase da edição de camadas. Podemos até ficar olhando pelos óculos enquanto tentamos ajustes. E podemos também forçar um pouco a realidade fazendo esse deslocamento lateral de maneira que os objetos venham para a frente ou para tras na cena. Para isto, novamente, o limite é a prática, e convém consultar pessoas com pouca familiaridade na visualização anaglífica para evitar que nossa vista, que começa a se acostumar com esforços pouco comuns, considere bom um resultado que a maioria não ia conseguir realizar.

**Rotações**

Se acontece que notamos alguns objetos na foto estarem acima ou embaixo dos demais na cena é porque houve elevação ou descida angular da câmera entre as duas tomadas, o que pode ser facilmente corrigido com o cursor de deslocamento do Gimp. Nesse caso de descontrole, porém, o mais comum é que aconteça uma rotação e alguns objetos da cena esquerda, por exemplo, estão por cima dos corresondentes na cena direita em uma parte da cena, e em outra parte da cena acontece o contário, e de maneira progressiva que acusa onde está o centro da rotação: alí onde um objeto coincide com seu correspondente. O comando de rotação do GIMP permite corrigir isto ("Transformações","Rotação"), vale a pena praticar porque depois poderá se aventurar e tirar as duas fotografias a partir de seu rosto, como em uma fotografia convencional. Primeiramente apoia bem o corpo nas duas pernas, prende a respiração e tira a primeira foto com a câmera enquanto olha enquadrando pelo olho esquerdo, e passa suavemente, conservando a cabeça no lugar o mais possível, a câmera para o olho direito. Quando tiver prática suficiente, poderá até fotografar pessoas, pedindo que fiquem quietas (melhor se sentadas ou apoiadas) entre a duas tomadas. Um objeto imóvel que justifica e muito a fotografia estéreo são as esculturas: Que sentido faz não tirar uma fotografia 3D de uma escultura, que é a arte onde não é usada a cor e sim a profundidade?

**Conflitos com cores**

A maneira acima indicada facilita resultados tridimensionalmente bons, mas sem reprodução de cores. O cérebro recebe impulsos vindos da cor vermelha no olho esquerdo e azul no direito, gerando a sensação de uma cor algo escura, que pode ser qualificada como marrom, totalmente artificial. Se deixarmos de anular o canal de cor verde na fotografia da vista direita, o cérebro consegue recompor algo parecido à tricromia original e o resultado pode ser muito bom e próximo do ideal. Recomendo para isto, de todas maneiras, o teste de olhar por um olho só para ver se o filtro não deixa passar "fantasmas", restos da imagem que não deveria passar. Outra complicação nisto são as cores fortes nas tonalidades vermelho e azul: acaba acontecendo que são vistas por somente um dos olhos e o cérebro não consegue compor a imagem, pior ainda, estranha a falta dela em um dos olhos e fica em conflito. Se não podemos evitar essas cores, ainda fica em caso de conflito o recurso de converter a tonalidades de cinza uma ou as duas fotografias, antes da edição.

**Óculos para estereoscopía anaglífica (estéreo bicolor)**

O 3D anaglifo é muito difundido na internet pela tradição que existe em outros paises, Miles de fotos são publicadas de muitos filmes também (no YouTube 3D). Vemos com muita frecuencia na internet páginas dizendo que é fácil construir seu próprio óculos bicolor(21), o que não é verdade. Ou é uma exageração resultante de ter lido materias escritas no estrangeiro, ou vem de uma falta total de apreciação da qualidade de um resultado em 3D. O cérebro exige uma excelente correlação entre o que percebe entre os dois olhos, qualquer diferença gera conflito. Não se consegue distribuir um corante ou pintar com caneta de maneira uniforme sobre plástico transparente. Nem temos no celofane nacional a qualidade que se encontra no estrangeiro, sobretudo nas cores azul e verde. Somente acrílico, que é muito grosso, e acetato, que é importado mas vende-se no pais, tem cores uniformes e podem filtrar com qualidade. Mas ainda é preciso muitíssimo cuidado na seleção do material, é extremamente difícil encontrar. Vende-se as vezes pela internet o acetato ou os óculos prontos, mas os preços são três vezes maiores que os do material importado diretamente. Para uma pequena prática, pode-se comprar, porém, para um uso mais massivo que pretenda minimizar o orçamento, recomendo uma procura criteriosa ou consulta a especialista. E todo cuidado é pouco para conservar o óculos limpo, nunca se deve tocar nos filtros, a limpeza não resulta fácil nem perfeita.

Uma outra observação é respeito ao uso dos anaglifos com projetores multimídia: somente os tradicionais, que funcionam baseados em cristal líquido ("LCD"), servem. Os que vem com o sistema de microespelhos ("DLP") não separam corretamente as cores básicas e não servem.

**Condições necessária para melhor ver fotos e vídeos 3D bicolor**
(chamadas tecnicamente "anaglifos")
1) Ter os óculos vermelho azul ou vermelho-verde de boa qualidade e limpos. Papel celofane não serve.
2) Conferir de ter o filtro vermelho para o olho esquerdo.
3) Não ter a vista cansada ou irritada.
4) No monitor de micro ou TV, ou no projetor, ter a sala o mais escurecida possível e o monitor no brilho máximo.
5) Se a visão não resulta confortável, tentar se afastar da tela.
6) Fica bem experimentar de dar zoom na foto.

Mesmo com tudo perfeito, há uma percentagem importante de pessoas que não conseguem ver bem as imagens. Isto é normal, visto que a estereoscopia não reproduz a cena exatamente como ela aparece aos olhos.

**Maus exemplos**

Por falta de conhecimento, editoras publicam fotos 3D com tintas erradas que depreciam aos olhos do público o uso do anaglifo. Temos na Fig. 3 um exemplo.

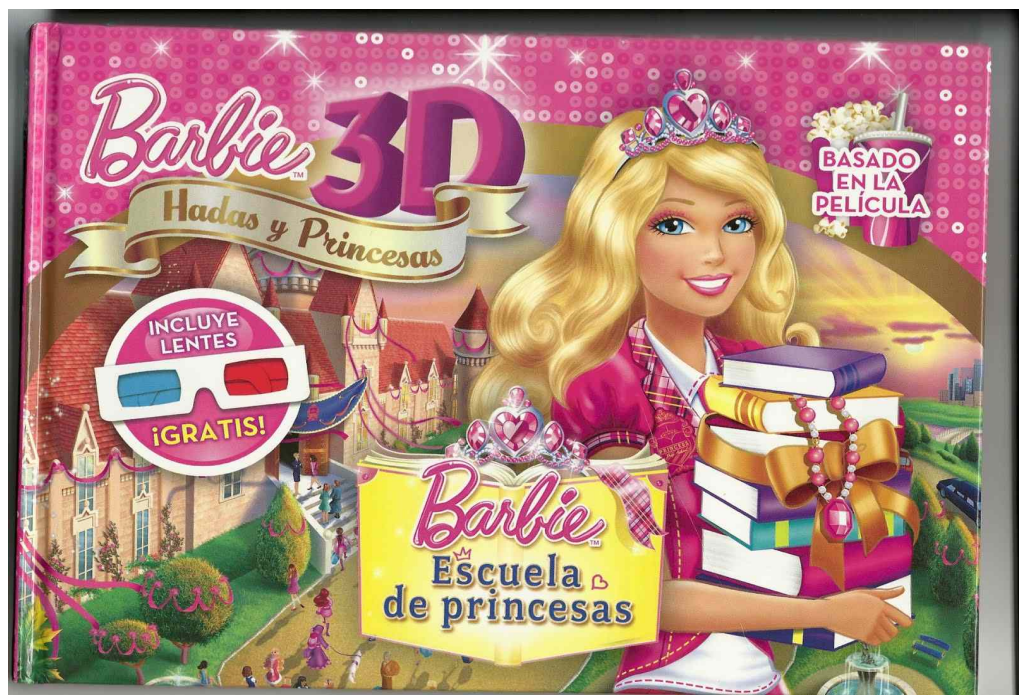

*Figura 3: Exemplo de livro realizado com má qualidade de impressão.*

**Montagem de fotos**

Esculturas são elementos que merecem ter suas fotos em 3D. Vejamos por exemplo o trabalho que tenho realizado recentemente com fotos da famosa escultura "La Pietá", de Michelangelo.

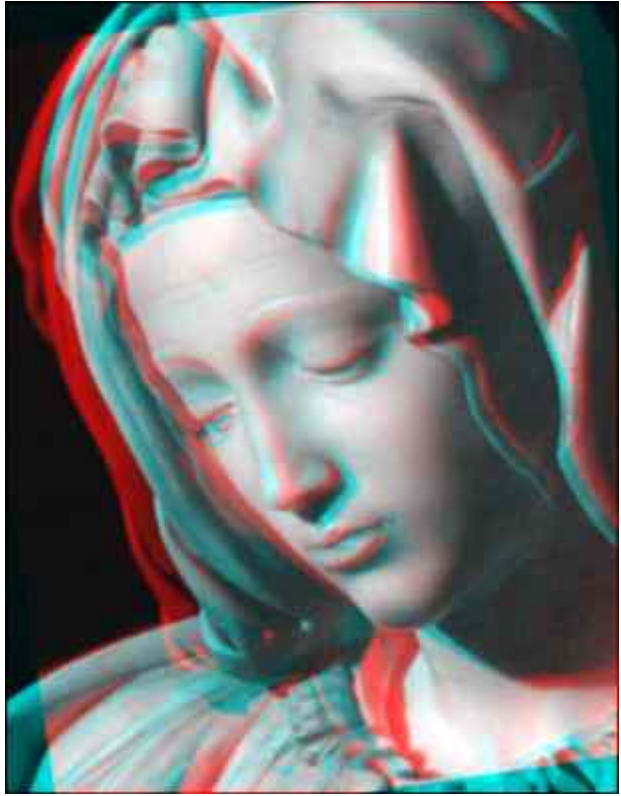

*Figura 3: exemplo de montagem de duas fotos não feitas especificamente para dar 3D. Autor: J.J. Lunazzi*

Somente posso atribuir à ignorância o fato de fotos de esculturas não serem feitas em 3D.

**Vídeo em 3D**

O processo para vídeo é semelhante, mas requer uma sincronização exata entre a pista que tem a vista esquerda e a que tem a direita. Essa é a parte difícil. Câmeras com relógios sincronizados e programas de edição profissionais devem facilitar a tarefa. De todas maneiras temos conseguido realizar vídeos usando o programa não livre SONY VEGAS Pro(14).

**Conclusões**

Descreve-se que, apesar de ser conhecida há muito tempo, uma técnica interessante no campo das imagens acabou sendo esquecida no Brasil e não veio a renascer sequer com a utilização ampla comercial no campo do entretenimento. Um fator que podemos assinalar é a falta de possibilidades de divulgação na mídia, sobretudo quando se quer acompanhá-la de elementos para o uso prático e não como a indicação de uma mercadoria a adquirir. O desconhecimento não é obstáculo para que seja aplicada como elemento didático e prático devido a que é possível praticá-la cada vez com mais facilidade por meio das tecnologias digitais de imagens e de processamento das mesmas. O engajamento das instituições que zelam pela divulgação científica seria o caminho para colocar o tema no nível que entendemos merece.

[20] Programa livre significa não apenas que é gratuito como que pode ser distribuido, copiado e usado livremente e, mais ainda, dá-se o código fonte para quem desejar conhecé-lo melhor ou modificá-lo. O Gimp está na versão 2.8 e obtem-se por [www.gimp.org](www.gimp.org)

[21] [http://www.3dblog.com.br/anaglifo/como-fazer-um-oculos-3d-anaglifo-passo-a-passo/](http://www.3dblog.com.br/anaglifo/como-fazer-um-oculos-3d-anaglifo-passo-a-passo/)

[22] José J. Lunazzi, evento de extensão oferecido na UNICAMP desde 1981.
[www.ifi.unicamp.br/~lunazzi/expo.htm](www.ifi.unicamp.br/~lunazzi/expo.htm)